\documentclass[10pt,twocolumn,letterpaper]{article}

\usepackage{cvpr}       

\usepackage{graphicx}
\usepackage{amsmath}
\usepackage{amssymb}
\usepackage{booktabs}
\usepackage{url}
\usepackage[accsupp]{axessibility}

\usepackage[pagebackref,breaklinks,colorlinks]{hyperref}

\newcommand\blfootnote[1]{%
	\begingroup
	\renewcommand\thefootnote{}\footnote{#1}%
	\addtocounter{footnote}{-1}%
	\endgroup
}

\usepackage[capitalize]{cleveref}
\crefname{section}{Sec.}{Secs.}
\Crefname{section}{Section}{Sections}
\Crefname{table}{Table}{Tables}
\crefname{table}{Tab.}{Tabs.}

\begin{document}

\title{The Devil Is in the Details: Window-based Attention for Image Compression}

\author{
    Renjie Zou $^{1}$ \hspace{10mm}
    Chunfeng Song $^{1}$ \hspace{10mm} 
    Zhaoxiang Zhang $^{1,2,*}$\\
    $^{1}$ National Laboratory of Pattern Recognition (NLPR) of\\
    Institute of Automation, Chinese Academy of Sciences (CASIA) \&\\
    University of Chinese Academy of Sciences (UCAS) \\
    $^{2}$ Centre for Artificial Intelligence and Robotics, HKISI\_CAS\\
{\tt\small \{zourenjie2020, chunfeng.song, zhaoxiang.zhang \}@ia.ac.cn}}

\maketitle
\blfootnote{* Corresponding Author}
\begin{abstract}

Learned image compression methods have exhibited superior rate-distortion performance than classical image compression standards. Most existing learned image compression models are based on Convolutional Neural Networks (CNNs). Despite great contributions, a main drawback of CNN based model is that its structure is not designed for capturing local redundancy, especially the non-repetitive textures, which severely affects the reconstruction quality. Therefore, how to make full use of both global structure and local texture becomes the core problem for learning-based image compression. Inspired by recent progresses of Vision Transformer (ViT) and Swin Transformer, we found that combining the local-aware attention mechanism with the global-related feature learning could meet the expectation in image compression. In this paper, we first extensively study the effects of multiple kinds of attention mechanisms for local features learning, then introduce a more straightforward yet effective window-based local attention block. The proposed window-based attention is very flexible which could work as a plug-and-play component to enhance CNN and Transformer models. Moreover, we propose a novel Symmetrical TransFormer (STF) framework with absolute transformer blocks in the down-sampling encoder and up-sampling decoder.
Extensive experimental evaluations have shown that the proposed method is effective and outperforms the state-of-the-art methods. The code is publicly available at \url{https://github.com/Googolxx/STF}. 
\end{abstract}

\begin{figure}[t]
\centering
\includegraphics[width=0.9\linewidth]{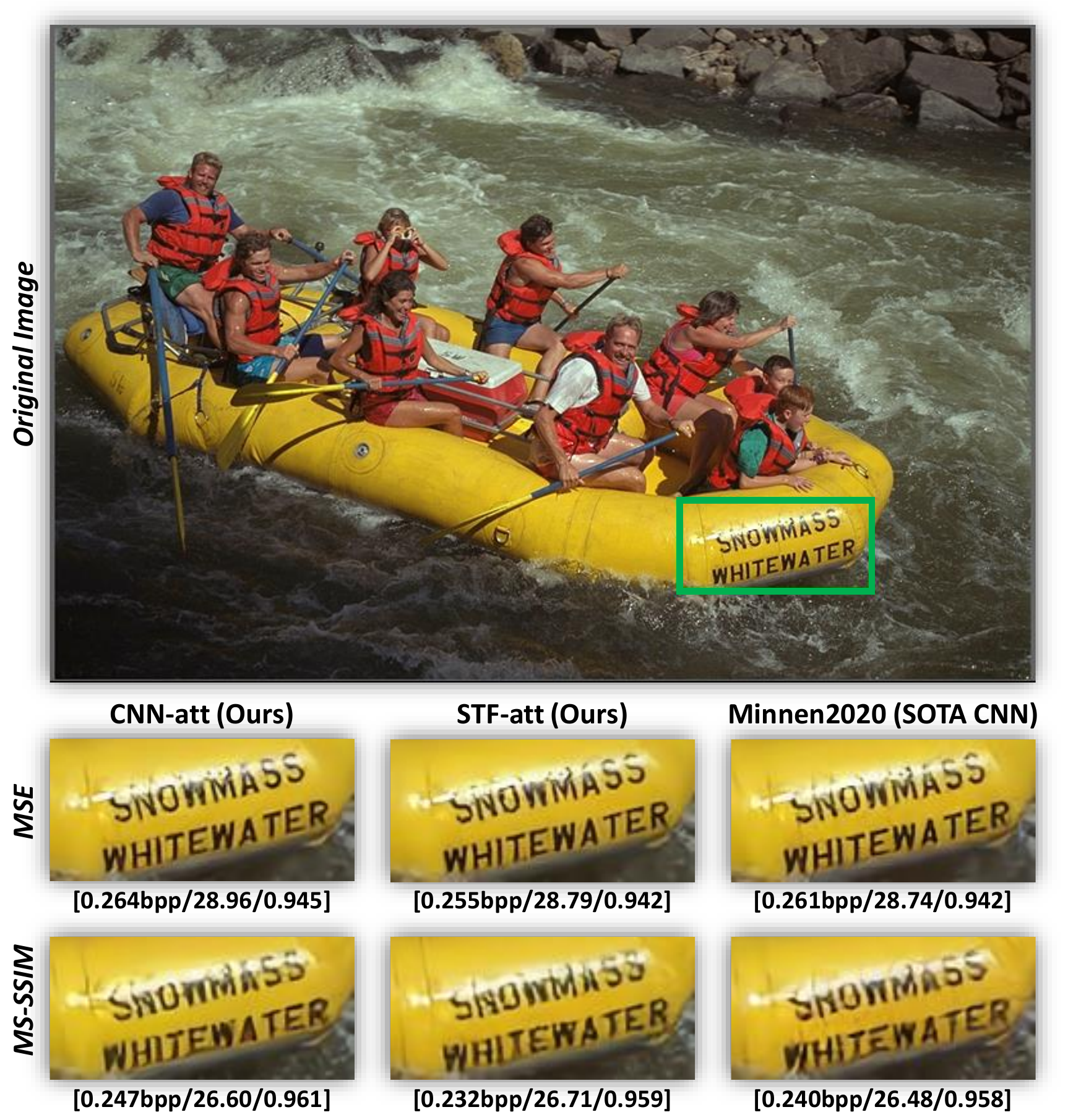}
\vspace{-1.0mm}
\caption{Comparison of image reconstruction by different methods, including CNN + window-attention (CNN-att), Symmetrical TransFormer + window-attention (STF-att), and previous SOTA CNN model (MBT).
The upper shows results optimized for MSE, while the lower is optimized for MS-SSIM.
It is obvious that the proposed window-based attention is effective for both supervision, especially with the STF model, showing that the proposed window-based attention model could internally allocate more bits on high contrast areas and achieve better visual quality.
The metrics are [bpp$\downarrow$/PNSR$\uparrow$/MS-SSIM$\uparrow$].}
\label{fig:attention contrast}
\vspace{-1.0mm}
\end{figure}

\section{Introduction}
\label{sec:intro}

Image compression is a fundamental and long-standing research topic in image processing field.
With the ever increasing visual applications, lossy image compression is a vital technique for storing images and videos efficiently in limited hardware resources.
Classical lossy image compression standards including JPEG \cite{wallace1992jpeg}, JPEG2000 \cite{rabbani2002jpeg2000}, BPG\cite{bpg}, and VVC \cite{vvc} follow a similar coding scheme: transforming, quantization, and entropy coding.
However, these image compression standards rely heavily on the hand-crafted rules, which means they are not expected to be of the ultimate solution for image compression.

In recent years, the learned image compression based on variational auto-encoder (VAE) \cite{2014Auto} has achieved better rate-distortion \cite{shannon2001mathematical} performance than conventional lossy image compression methods on metrics of Signal-to-Noise Ratio (PSNR) and Multi-Scale-Structural Similarity Index Measure (MS-SSIM) \cite{wang2003multiscale}, showing great potential for practical compression use. Here, we would briefly introduce the general pipeline \cite{balle2017end} of the VAE-based methods. For encoding, the VAE-based image compression methods use a linear and nonlinear parametric analysis transform to map the images to a latent code space. After quantization, entropy estimation modules predict the distributions of latents, then the lossless Context-based Adaptive Binary Arithmetic Coding (CABAC) \cite{marpe2003context} or Range 
Coder \cite{martin1979range} compresses the latents into the bit stream. Meanwhile, hyper-prior \cite{balle2018variational}, auto-regressive \cite{minnen2018joint} priors and Gaussian Mixture Model (GMM) \cite{cheng2020learned} allow the entropy estimation modules to more precisely predict distributions of latents, and achieve better rate-distortion (RD) performance. For decoding, lossless CABAC or Range Coder decompresses the bit stream, then the decompressed latents are mapped to reconstructed images by a linear and nonlinear parametric synthesis transform. Further, there are also some works \cite{lee2019end, hu2018combine, zhou2018variational} designing post-processing networks for better quality of reconstruction. Combining above sequential units, those models could be trained end-to-end. Although great progresses have been achieved, one core problem of above CNN-based model is that the original convolutional layer is designed for the high-level global feature distillation, rather than the low-level local detail restoration. As shown in the right side of Fig.\ref{fig:attention contrast}, even the SOTA CNN model is still affected by the weak local detail learning ability which would inevitably limit further performance improvement. 

Inspired by the success of attention mechanism in natural language processing (NLP) and computer vision tasks such as the image classification and semantic segmentation, many researchers apply the non-local attention mechanism to guide the adaptive processing of latent features, which could help the compression algorithm allocate more bits to challenging areas (i.e., edges, textures) for better RD performance. However, such non-local attentions still not change the intrinsic global-aware character of the CNN structure. Recent studies \cite{carion2020end, dosovitskiy2020image, Liu_2021_ICCV, touvron2021training} have demonstrated that transformer \cite{vaswani2017attention} can be successfully applied to vision tasks with competitive even better performance compared with convolutional neural networks (CNNs). Those attention-based networks, such as Vision Transformer \cite{dosovitskiy2020image} and Swin Transformer \cite{Liu_2021_ICCV} take the advantages of attention mechanism to capture global dependency. However, we intuitively find that the global semantic information in image compression is not as effective as in other computer vision tasks. Instead, spatially neighboring elements have stronger correlation.

Following above discussions, this paper explores to address the detail missing problem from two aspects, i.e., studying \emph{the local-aware attention mechanism} and introducing \emph{a novel transformer-based framework}. \textbf{First}, we comprehensively study how to combine the neural networks with the attention mechanism for designing local-aware lossy image compression architectures. Through conducting a set of comparative experiments based on a global attention mechanism and a local attention mechanism, we have verified our aforementioned guess, that the local attention is more suitable for the local texture reconstruction. Afterward, we present a flexible attention module combined with neural networks to capture correlations among spatially neighboring elements, namely window-attention. As shown in Fig.\ref{fig:attention contrast}, the proposed attention module could work as a plug-and-play component to enhance CNN and Transformer models. \textbf{Second}, although the transformer-based models have achieved great success in a variety of computer vision tasks, there are still great challenges applying transformer model in image compression, e.g., no up-sampling units, fixed attention model. To this end, we propose a novel Symmetrical TransFormer (STF) framework with absolute transformer blocks in down-sampling encoder and up-sampling decoder, which may be the first exploration of designing up-sampling transformer, especially for the image compression task. Extensive experimental results show that our methods are superior to the state-of-the-art (SOTA) image compression methods in essential metrics. The main contributions of this paper are summarized as follows:
\vspace{-1.0mm}
\begin{itemize}
\setlength{\itemsep}{0pt}
\setlength{\parsep}{0pt}
\setlength{\parskip}{0pt}
    \item We extensively study the local-aware attention mechanism, and find that it is crucial to combine the global structure learned by neural networks and the local texture mined by the attention units.
    \item We present a flexible window-based attention module to capture correlations among spatially neighboring elements, which could work as a plug-and-play component to enhance CNN or Transformer models.
    \item We design a novel Symmetrical TransFormer (STF) framework with absolute transformer blocks in both down-sampling encoder and up-sampling decoder.
    \item Extensive experimental evaluations have shown that the proposed methods are effective and outperform the SOTA image compression methods.
\end{itemize}

\section{Related Works}
\noindent\textbf{Learned Image Compression.} Recently, learned image compression models \cite{balle2017end, balle2018variational, minnen2018joint, lee2019end, cheng2020learned, minnen2020channel, cui2021asymmetric, he2021checkerboard, choi2019variable, mentzer2018conditional} based on CNNs exhibit a fast development trend and achieve significant breakthroughs.
For VAE-based architectures, \cite{balle2017end} firstly proposes an end-to-end optimized model for image compression.
\cite{balle2018variational} incorporates a hyper-prior to effectively capture spatial dependencies in the latent representation based on \cite{balle2017end}.
Inspired by the success of auto-regressive priors in probabilistic generative models, \cite{minnen2018joint} further enhances the entropy model in \cite{balle2018variational} by adding an auto-regressive component.
In addition to this, \cite{cheng2020learned} enhances the network architecture by using residual blocks and adding a simplified attention module with the Gaussian Mixture Model (GMM) replacing the mostly used Single Gaussian Model (SGM).
Though the GMM-based entropy model performs better in RD performance, the adopted probability distribution function (PDF) and cumulative distribution function (CDF) have to be dynamically generated for every single element in encoding and decoding, thereby introducing much redundancy and making it time-consuming.
In contrast, the SGM-based entropy model can build fixed PDF and CDF tables for entropy coding, which is less computationally expensive.
To minimize serial processing steps in auto-regressive context models, \cite{minnen2020channel} proposes a channel-wise auto-regressive entropy model.
Some approaches \cite{agustsson2019generative, mentzer2020high} use the Generative Adversarial Networks (GANs) to directly learn the distribution of images and prevent compression artifacts.
For GAN-based architectures, image compression is a rate-distortion-perception trade-off task.

\noindent\textbf{Attention Mechanism.} Attention mechanism mimics the internal process of biological observation, devoting more attention resources to the key regions to obtain more details and suppress other useless information.
Non-local attention mechanism \cite{wang2018non} has been proven beneficial in various vision tasks.
For learned image compression, \cite{liu2019non} applies the non-local attention to generate implicit importance masks to guide the adaptive processing of latent features, while \cite{cheng2020learned} simplifies the attention mechanism by removing the non-local block.

\noindent\textbf{Transformer-based Model.} Inspired by the success of Transformer architectures \cite{vaswani2017attention} in the natural language processing, there are many works exploring the potentiality of Transformer in computer vision tasks. \cite{dosovitskiy2020image} conducts image classification with transformer architecture. \cite{carion2020end} implements transformer-based models in detection. \cite{chen2021pre} proposes a universal pre-training approach for image processing tasks.
\cite{Liu_2021_ICCV} proposes a hierarchical Transformer computed with shifted windows and achieves SOTA performance in image classification, semantic segmentation, and object detection.
However, as far as our knowledge, there is no transformer related work for image compression.
In this paper, we explore how to design a transformer-based architecture for learned image compression to achieve comparable RD performance.

\noindent\textbf{Normalization in Neural Networks.} Generalized divisive normalization (GDN) \cite{balle2016density} is a milestone in learned image compression.
GDN is highly efficient in Gaussianizing the local joint statistics of natural images.
We have conducted some comparative experiments by replacing GDN with batch normalization \cite{ioffe2015batch}, layer normalization \cite{ba2016layer}, and channel normalization \cite{mentzer2020high}.
We found that replacing GDN in CNN would result in a huge drop in RD performance.
However, we also found that GDN is unstable in a deep Transformer-based architecture.
In addition, GDN is not compatible with the attention mechanism in Transformer blocks.

\section{Method}
\subsection{Formulation}
Since our method is built upon the hyper-prior architecture \cite{balle2018variational, minnen2018joint} and channel-wise auto-regressive entropy model \cite{minnen2020channel}, we would briefly introduce the basic pipeline for better understanding.

The encoder $E$ maps a given image $x$ to a latent $y$. After quantization $Q$,
$\hat{y}$ is the discrete representation of the latent $y$.
Then $\hat{y}$ is mapped back to the reconstructed image $\hat{x}$ using the decoder $D$.
The main process is formulated as:
\begin{equation}
    \begin{split}
        y &= E(x;\phi)\\
        \hat{y} &= Q(y) \\
        \hat{x} &= D(\hat{y};\theta)
    \end{split}
\end{equation}
where $\phi$ and $\theta$ are trainable parameters of the encoder $E$ and decoder $D$.
Quantization $Q$ inevitably introduces clipping errors of the latent ($error = y - Q(y)$), which leads to distortion of the reconstructed image.
Following previous work\cite{minnen2020channel}, in the training phase, we also modify the quantization error by rounding and adding the predicted quantization error.

We model each element $\hat{y}_{i}$ as a single Gaussian distribution with its standard deviation $\sigma_{i}$ and mean $\mu_{i}$ by introducing a side information $\hat{z}_{i}$.
The distribution $p_{\hat{y}_{i}|\hat{z}_{i}}$ of $\hat{y}_{i}$ is modeled by a SGM-based entropy model:
\begin{equation}
    \begin{split}
        p_{\hat{y}_{i}|\hat{z}_{i}} (\hat{y}_{i}|\hat{z}_{i}) = \mathcal{N}(\mu_{i},\sigma_{i}^{2})
    \end{split}
\end{equation}
The loss function of image compression model is:
\begin{equation}
    \begin{split}
        \mathcal{L} = &R + \lambda\cdot D \\
        = &\mathbb{E}_{x\sim p_{x}}[-\log_{2} p_{\hat{y}|\hat{z}}({\hat{y}|\hat{z}}) - \log_{2}p_{\hat{z}} (\hat{z})] \\
        &   + \lambda \cdot \mathbb{E}_{x\sim p_{x}}[d(x, \hat{x})]
    \end{split}
\end{equation}
where $\lambda$ controls the trade-off between rate and distortion, $R$ is the bit rate of latents $\hat{y}$ and $\hat{z}$, $d(x, \hat{x})$ is the distortion between the raw image $x$ and reconstructed image $\hat{x}$.

\subsection{Window-based Attention}
Most of previous approaches \cite{liu2019non, zhou2019end, cheng2020learned} apply the attention mechanism to generate attention masks based on a global receptive field.
Attention mechanism has also been successfully adopted in many computer vision tasks, such as the image classification, semantic segmentation and object detection.
However, intuitively, global semantic information in image compression is not as practical as in those computer vision tasks.

\begin{figure}[!t]
\begin{center}
\includegraphics[width=0.95\linewidth]{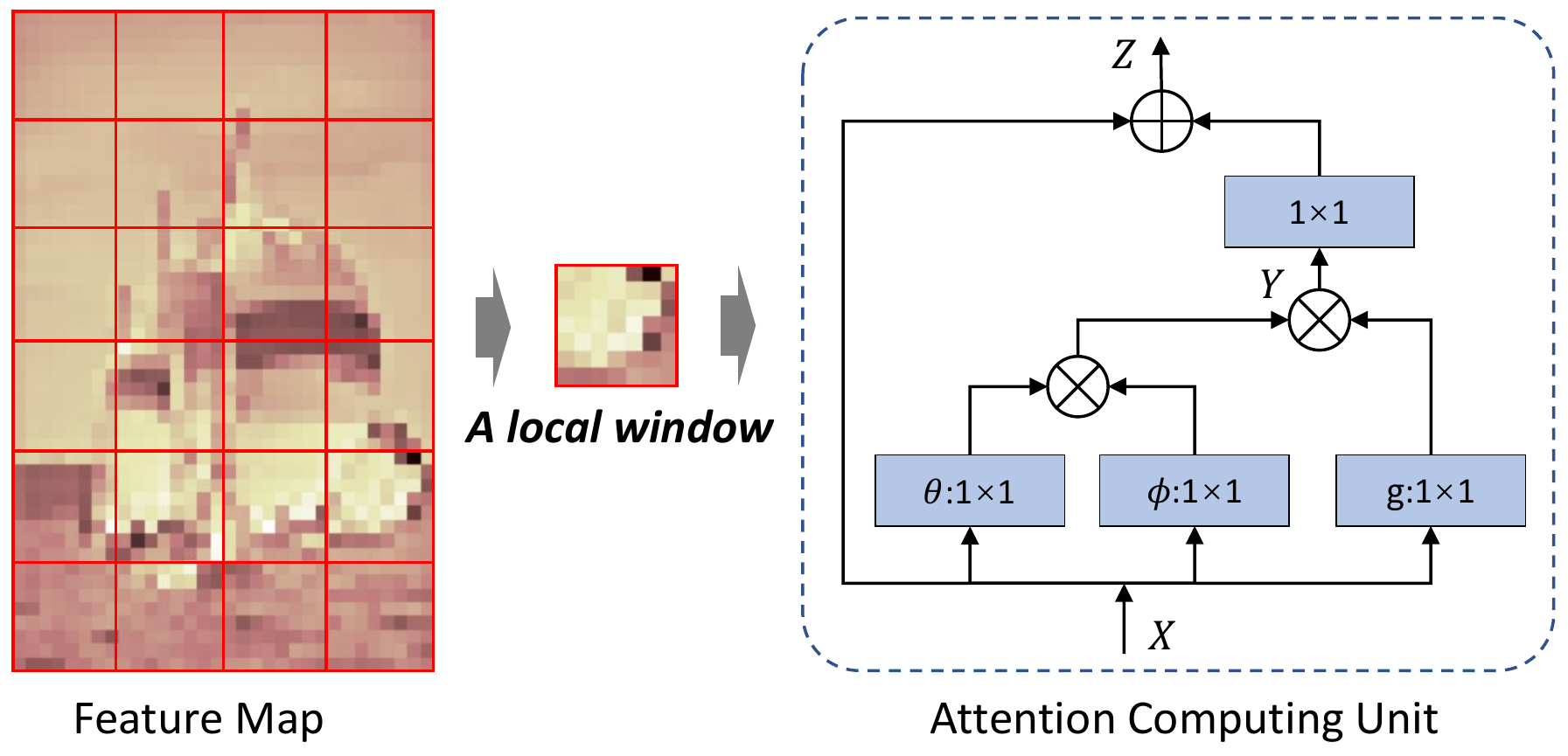}
\end{center}
\vspace{-1.0mm}
\caption{An illustration of the window-based attention. Attention masks are computed in a local window.}
\label{fig:window attention}
\vspace{-1.0mm}
\end{figure}

\noindent\textbf{Attention in Non-overlapped Windows.} We notice that generating attention masks based on spatially neighboring elements can improve RD performance with less computing cost.
For efficient modeling and focusing on spatially neighboring elements, we propose the window-based attention.
As illustrated in Fig. \ref{fig:window attention}, the feature map is divided into windows of $M\times M$ in a non-overlapping manner.
We compute the attention map in each window separately, $X_{i}^{k}$ and $X_{j}^{k}$ are $i$-th and $j$-th elements in $k$-th window, as below:
\begin{equation}
    \begin{split}
        Y_{i}^{k} = \frac{1}{C(X^{k})} \sum_{\forall j} f(X_{i}^{k},X_{j}^{k})g(X_{j}^{k})
   \end{split}
   \label{eq:y_k}
\end{equation}
\vspace{-2.0mm}
\begin{equation}
    \begin{split}
        \text{\textbf{with  }} f(X_{i}^{k},X_{j}^{k}) &= e^{\theta(X_{i}^{k})^{T} \phi(X_{j}^{k})},\\
        C(X^{k}) &= \sum_{\forall j}f(X_{i}^{k},X_{j}^{k}), \\
        g(X_{j}^{k}) &= W_{g}X_{j}^{k} \\
   \end{split}
\end{equation}
Here, $\theta(X_{i}^{k}) = W_{\theta}X$ and $\phi(X_{i}^{k}) = W_{\phi}X$, where $W_{\theta}$ and $W_{\phi}$ are cross-channel transforms.
$f(\cdot)$ is an embedded Gaussian function.
$C^{k}(X)$ is a normalizing factor.
For given $i$ and $k$, $\frac{1}{C^{k}(X)} f(X_{i}^{k},X_{j}^{k})$ is the $softmax$ computation along the dimension $j$ in $k$-th window.
Residual connection is necessary for such attention mechanism, the output is as follows:
\begin{equation}
    \begin{split}
     Z_{i}^{k} = W_{z}Y_{i}^{k} + X_{i}^{k}
    \end{split}
\end{equation}
where $W_{z}$ is a weight matrix for computing the
position-wise embedding on $Y_{i}^{k}$, given in Eq.\ref{eq:y_k}.

\begin{figure}[!t]
\centering
\includegraphics[width=0.95\linewidth]{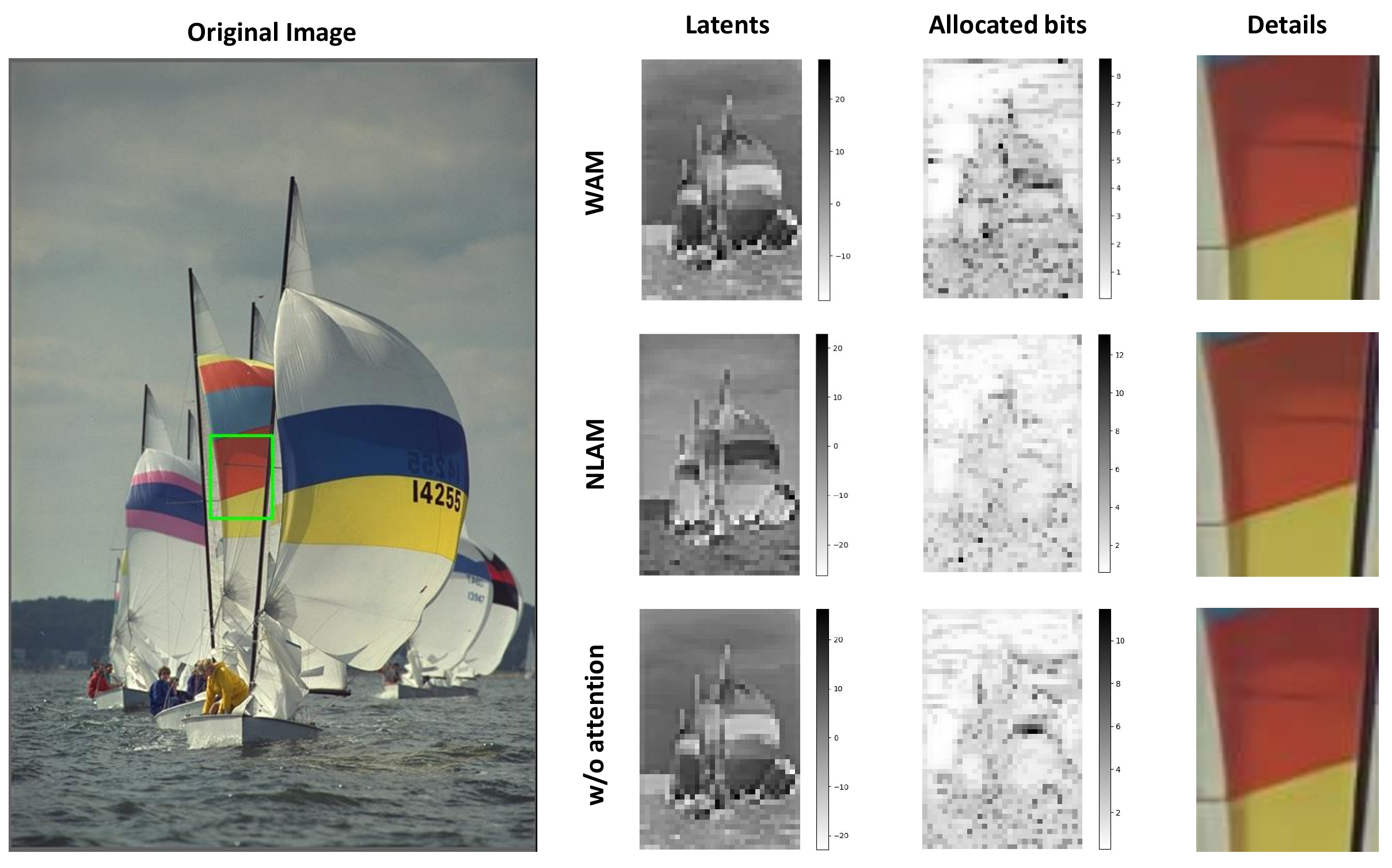}
\caption{Visualization of WAM, NLAM, and w/o attention module for the channel with maximal entropy.
It shows that our WAM focuses on high contrast regions (sailboats) and allocates more bits on those areas while fewer bits on low contrast regions (sky and clouds).
Instead, the bits are allocated evenly in the w/o attention module and NLAM.}
\label{fig:attention}
\vspace{-2.0mm}
\end{figure}

\begin{figure*}[!t]
\centering
\includegraphics[width=0.85\linewidth]{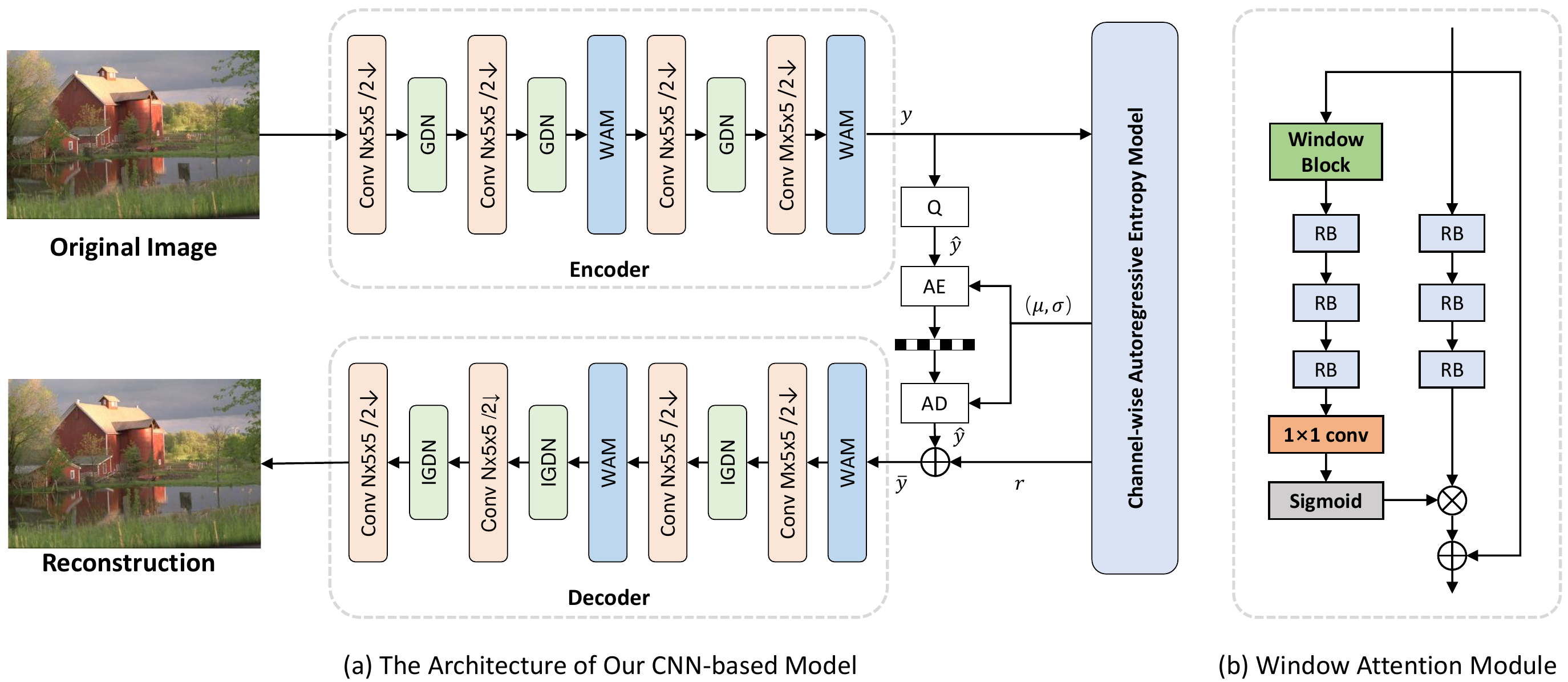}
\vspace{-1.0mm}
\caption{(a) The architecture of our proposed CNN-based model. We adopt the architecture of \cite{minnen2020channel} for the sake of subsequent comparison. IGDN is the inverse GDN. (b) RB is the residual block which consists of $1\times1$ and $3\times3$ convolutional layers.}
\label{fig:cnn arch}
\vspace{-1.0mm}
\end{figure*}

\noindent\textbf{Window Attention Module.} Liu et al. \cite{liu2019non} propose the Non-Local Attention Module (NLAM), which consists of a cascade of a Non-local Block and regular convolutional layers.
We replace the Non-local Block with Window Block for focusing on regions with high contrast.
Fig. \ref{fig:cnn arch} (b) shows our Window Attention Module (WAM).
We visualize the channel with the highest entropy for WAM, NLAM, and w/o attention module, as shown in Fig. \ref{fig:attention}.
In the 3-rd column, it's obviously observed that WAM can allocate more bits in complex regions (high contrast) and fewer bits in simple regions (low contrast).
In addition, the WAM reconstructed image has sharper texture details.
NLAM with global receptive field often leads to evenly allocation of bits on different regions, which is not consistent with the expectation in \cite{liu2019non}.

\subsection{CNN-based Architecture}
As shown in Fig. \ref{fig:cnn arch}, our CNN-based architecture is established upon \cite{minnen2020channel}.
We enhance the encoder and decoder by inserting the proposed WAMs, respectively.
The WAMs help allocate bits on different areas more rationally and internally with negligible computation overhead.
It is simply yet could significantly improve RD performance.

\subsection{Transformer-based Architecture}
\begin{figure*}[!t]
\centering
\includegraphics[width=0.85\linewidth]{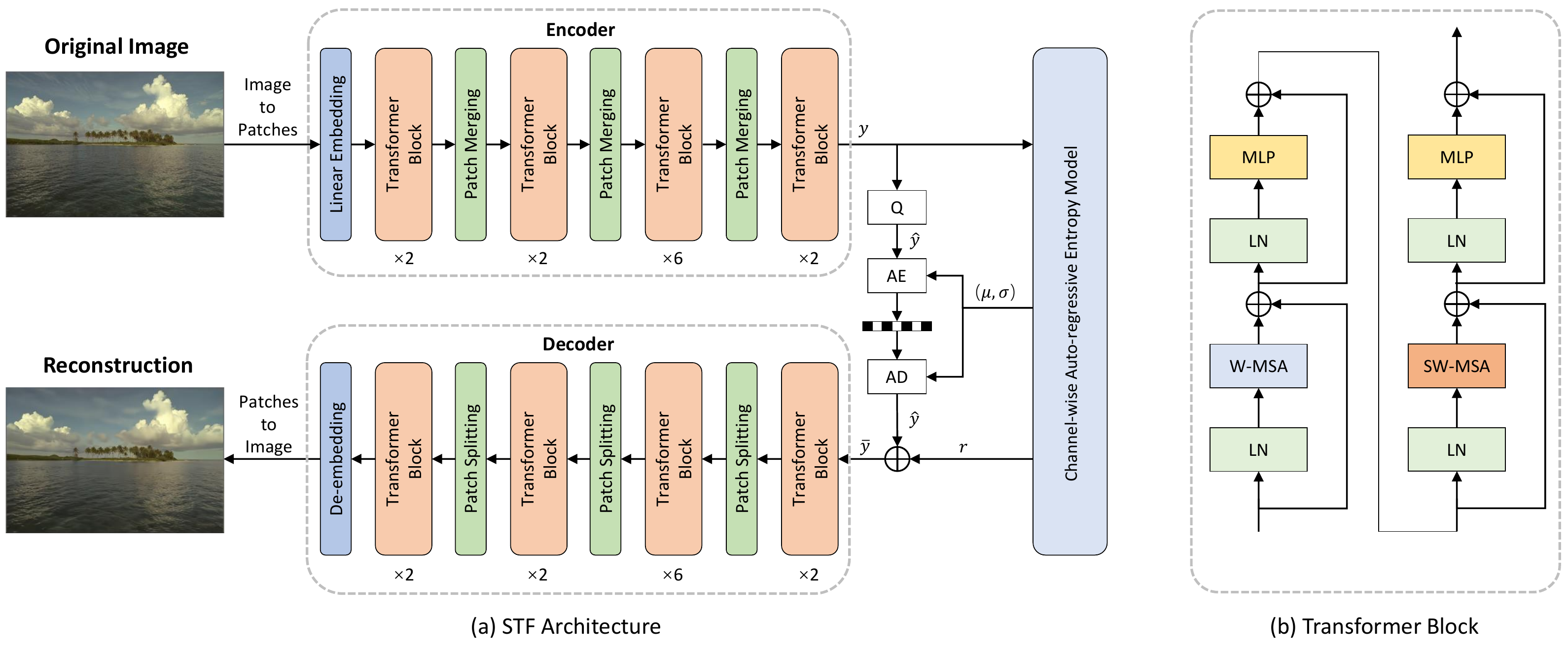}
\vspace{-1.0mm}
\caption{(a) The architecture of our proposed Transformer-based model. The patch merging layer and patch splitting layer consist of linear layers and LN layers. (b) W-MSA \cite{Liu_2021_ICCV} is the window-based multi-head self-attention, and SW-MSA \cite{Liu_2021_ICCV} is the shifted window-based self-attention, respectively.}
\label{fig:STF}
\vspace{-1.0mm}
\end{figure*}

Inspired by the success of Transformer architectures\cite{dosovitskiy2020image, Liu_2021_ICCV} in computer vision and our aforementioned experimental result that the local attention helps allocate bits rationally and improve RD performance, we further propose a novel Transformer architecture for learned image compression, as shown in Fig. \ref{fig:STF}.

\noindent \textbf{Rethinking the Transformer.} Since our goal is to verify whether combining self-attention layers and MLPs can achieve comparable performance with original CNN-based architectures, we design a novel Symmetrical Transformer (STF) framework without convolution layers in the encoder and decoder.
The difficulties of designing the Transformer model for learned image compression are the following:
\vspace{-1.0mm}
\begin{itemize}
\setlength{\itemsep}{0pt}
\setlength{\parsep}{0pt}
\setlength{\parskip}{0pt}
    \item Previous works are mostly based on the CNNs to eliminate the spatial redundancy and capture the spatial structure.
    Directly splitting image to patches may result in space redundancy within each patch.
    \item GDN is the most commonly used normalization and nonlinear activation function in image compression.
    However, GDN is unstable in the deep Transformer-based architecture.
    Furthermore, we have found GDN and the attention mechanism in the transformer are incompatible.
    \item Based on our previous analysis and experimental results, calculating the attention map on a large field is not optimal.
\end{itemize}
\vspace{-1.0mm}
To address above concerns, we choose a small patch size to avoid the space redundancy within each patch.
We use the LN for normalization, which is most common used in Transformer.
GELU is adopted as the nonlinear activation function in our Transformer architecture.
Inspired by \cite{Liu_2021_ICCV}, we compute the attention map within local windows.
The advantage of our Transformer architecture is that it could focus on spatially neighboring patches while gradually expanding the receptive field, with acceptable computational complexity.

\noindent \textbf{Transformer-based Encoder.} We split the raw image $x\in R^{3\times H\times W}$ into patches with patch size $N$.
A linear embedding layer is applied on the raw patches to generate a feature map $f_{p}\in R^{C \times \frac{H}{N}\times \frac{W}{N}}$ with $C$ channels.
The feature map $f_{p}\in R^{C \times \frac{H}{N}\times \frac{W}{N}}$ is reshaped into a sequence $f_{s}\in R^{P^{2}\times C}$, where $P = \frac{HW}{N^{2}}$ is the number of patches.
Then the sequence $f_{s}$ will be input into Transformer blocks and patch merging layers.
Following the structure in \cite{Liu_2021_ICCV}, the former calculate attention masks in a window for feed-forwarding.
At the same time, the latter down-sample the resolution of features and doubles channels of features.

\noindent \textbf{Transformer-based Decoder.} We design a symmetrical decoder consisting of multiple Transformer blocks, patch splitting layers and a de-embedding layer.
The patch splitting layers up-sample the resolution of features and halve channels of features.
The de-embedding layer maps the feature map to the reconstructed image $\hat{x}$.

\noindent \textbf{Entropy Model.} To more effectively and efficiently predict the probability distribution of the latent.
We use an SGM-based channel-wise auto-regressive entropy model \cite{minnen2020channel}.

\section{Experiments}
\subsection{Experimental Setup}
\textbf{Training.} We implement our proposed CNN-based architecture and Transformer-based architecture in CompressAI platform \cite{begaint2020compressai}. For training, we randomly choose 300k images from the OpenImages dataset \cite{openimages}, and randomly crop them with the size of $256 \times 256$.
All models are trained for 1.8M steps using the Adam optimizer \cite{kingma2015adam} with a batch size of 16. The initial learning rate is set to $1\times 10^{-4}$ for $120k$ iterations, and drops to $3\times 10^{-5}$ for another $30k$ iterations, $1\times 10^{-5}$ for the last $30k$ iterations.

Our models are optimized using two quality metrics (MSE and MS-SSIM) as supervisions.
Following the same settings in \cite{begaint2020compressai}, when the model is optimized for MSE, the lambda values $\lambda$ belongs to \{0.0018, 0.0035, 0.0067, 0.0130, 0.025, 0.0483\}.
When the model is optimized for MS-SSIM, the lambda values $\lambda$ belongs to \{2.4, 4.58, 8.73, 16.64, 31.73, 60.50\}.

For our CNN-based models, the channel number of latent and hyper latent are set to 320 and 192, respectively.
For our Transformer-based models, patch size is  $2$, window size is $4$, and channel number $C$ is $48$.
Typically, our models have same hyper-parameters for different $\lambda$.

\textbf{Evaluation.}
We evaluate our CNN-based models and STF models by calculating the average RD performance (PSNR and MS-SSIM) on the commonly used Kodak image set \cite{kodak} and CLIC professional validation dataset \cite{clic}. 
We compare our methods with influential learned compression methods, including the context-free hyperprior model (Ball\'{e}2018) \cite{balle2018variational}, auto-regressive hyper-prior model (Minnen2018) \cite{minnen2018joint}, and auto-regressive hyper-prior model with GMM and simplified attention (Cheng2020) \cite{cheng2020learned}.
Refer to appendix for RD curves covering a wide range of conventional and ANN-based compression methods.

\begin{figure*}[!t]
\centering
\includegraphics[scale=0.45]{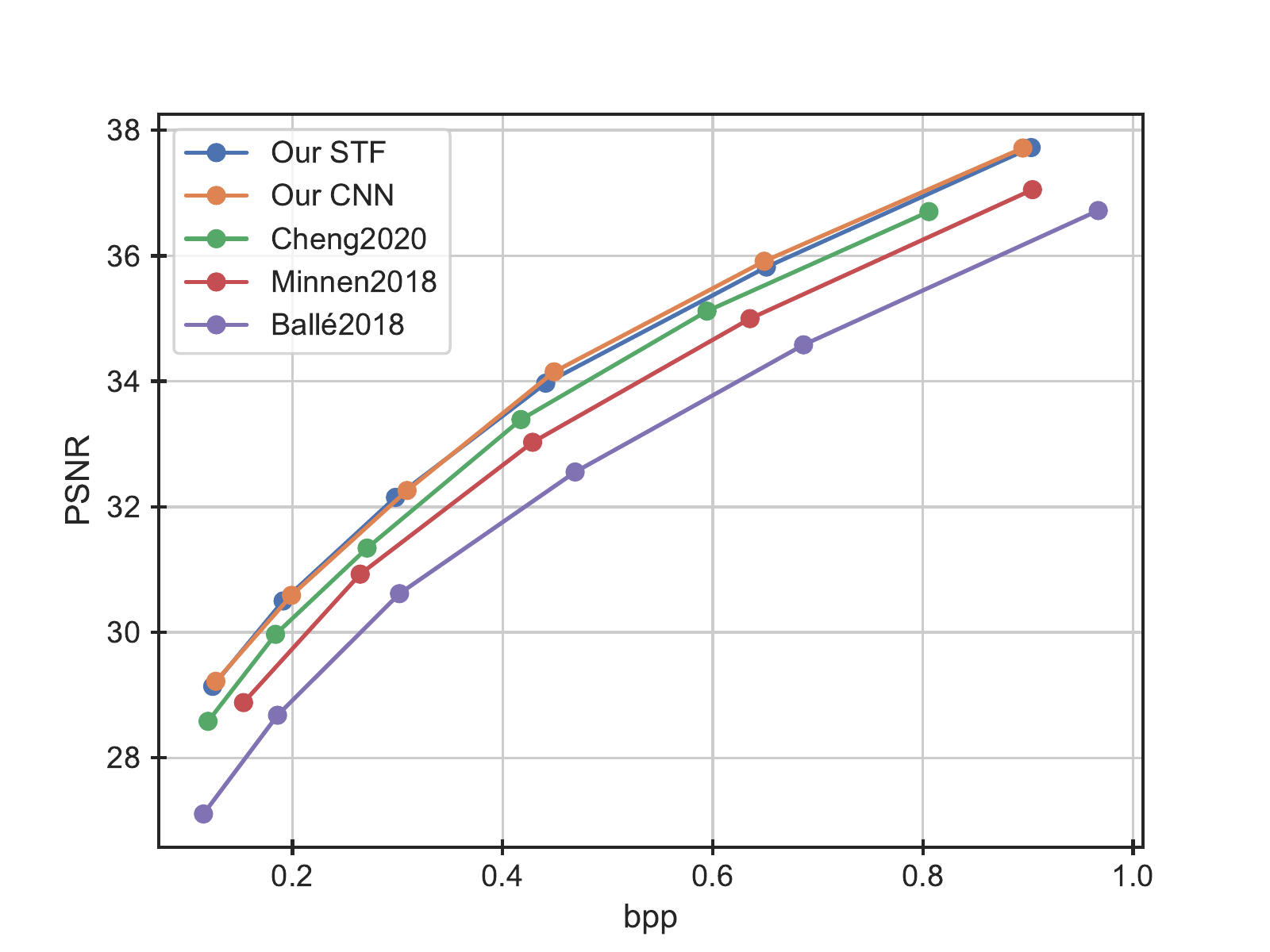}
\hspace{0.3in}
\includegraphics[scale=0.45]{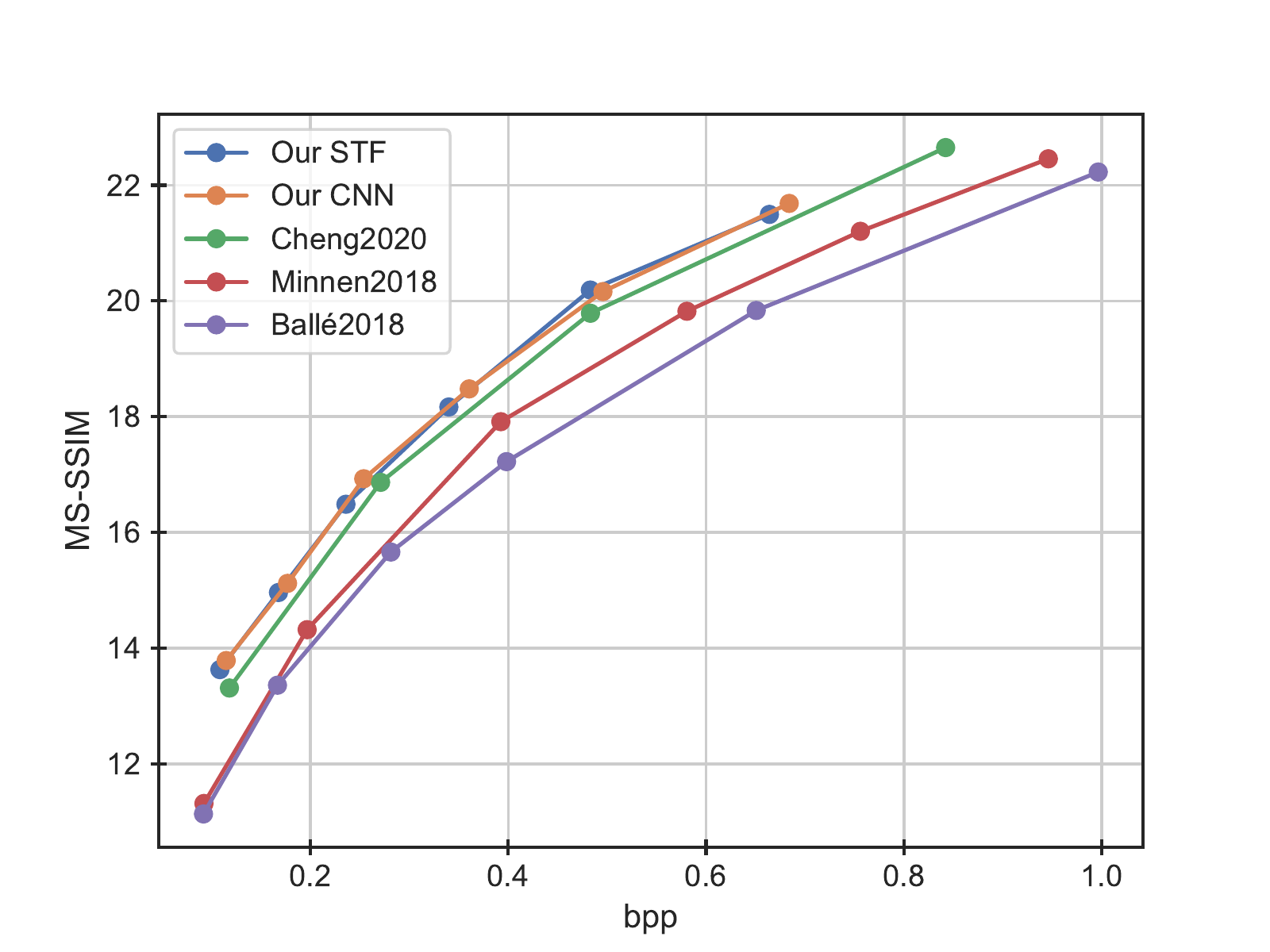}

\caption{RD Performance evaluation on the Kodak dataset, which contains 24 high quality images.}
\label{fig:kodak_rd}
\vspace{-2.0mm}
\end{figure*}

\begin{figure*}[!t]
\centering
\includegraphics[scale=0.45]{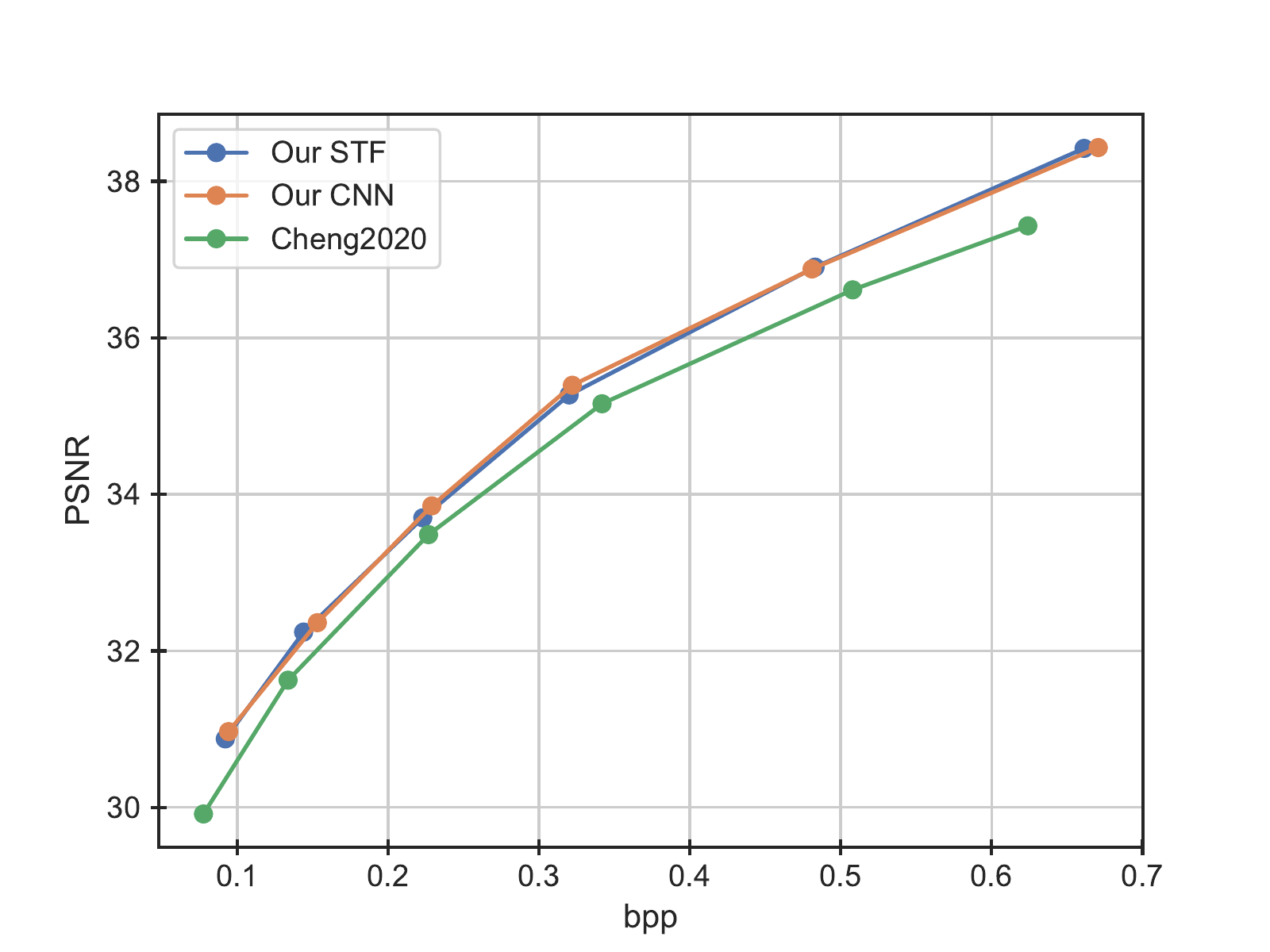}
\hspace{0.3in}
\includegraphics[scale=0.45]{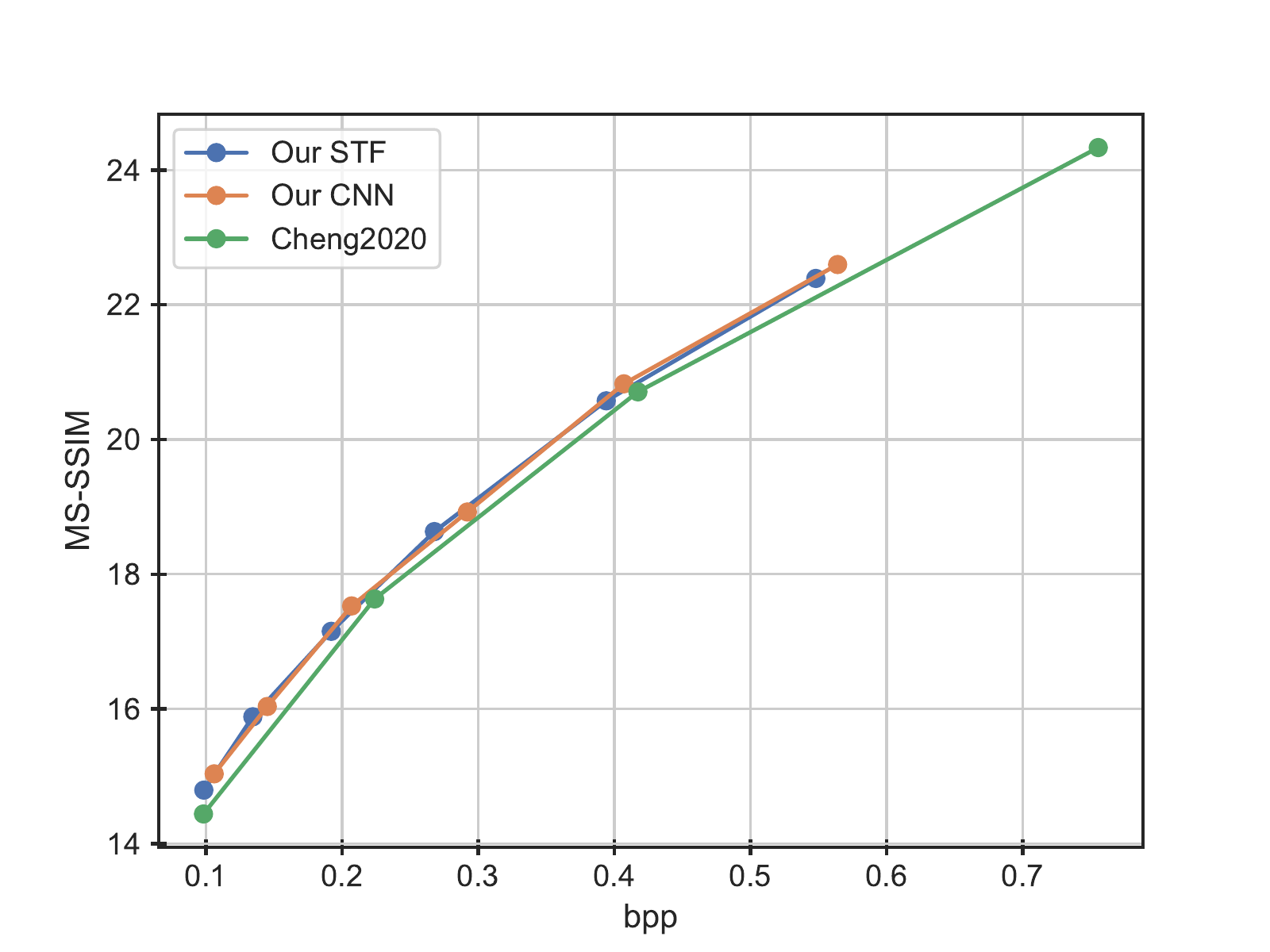}
\caption{RD Performance evaluation on CLIC Professional Validation dataset, which contains 41 high resolution and high quality images.}
\label{fig:clic_rd}
\vspace{-2.0mm}
\end{figure*}

\subsection{Comparison with the SOTA Methods}
\textbf{RD Performance.}
Fig. \ref{fig:kodak_rd} shows the comparison results on Kodak dataset.
When trained for MSE and measured by PSNR, the performances of our CNN-based models and STF models are very close, and could outperform other learned compression methods.
In contrast, when trained and measured by MS-SSIM, our CNN-based models and STF models only have slight improvements.
As mentioned in \cite{balle2018variational}, MS-SSIM has the effect of attenuating the error in regions with high contrast, and boosting the error in regions with low contrast.
But the fact is not as expected, it frequently assigns more details to the regions with low contrast (e.g., grass and hair) and removes details from regions with high contrast (e.g., text and salient objects).
Attention mechanism focuses more on high contrast regions, and consequently allocates more bits on them.
The contradiction may result in an inconspicuous improvement for our attention-based models when optimized for MS-SSIM.

As illustrated in Fig. \ref{fig:clic_rd}, the comparison results on the CLIC professional validation dataset indicate the same conclusion.
It shows the robustness of our CNN-based models and STF models.

\begin{figure*}[!t]
\vspace{-2.0mm}
\centering
\includegraphics[width=0.90\linewidth]{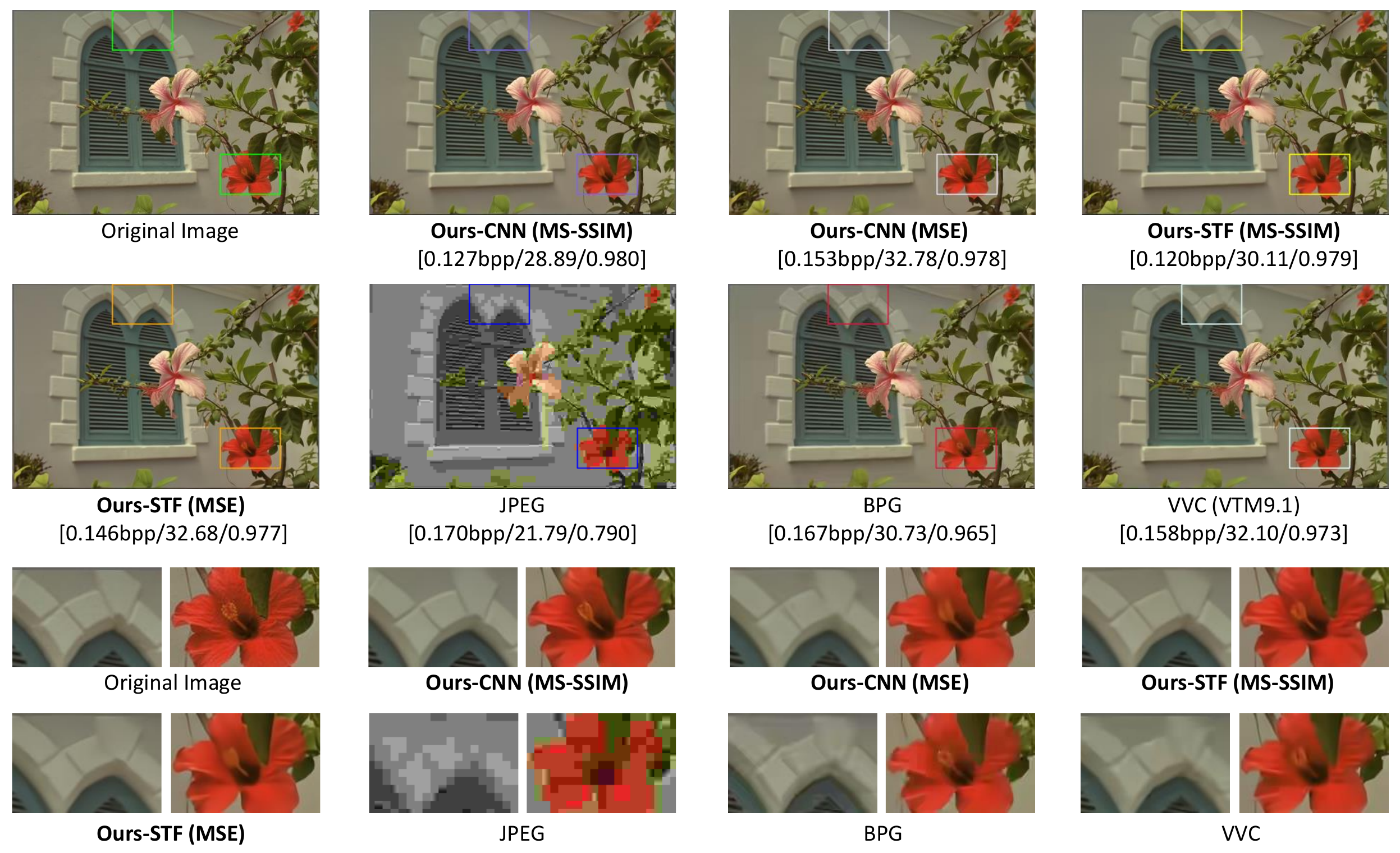}
\caption{Visualization of the reconstructed image (\emph{kodim07.png}) from Kodak dataset.
The metrics are [bpp$\downarrow$/PNSR$\uparrow$/MS-SSIM$\uparrow$].}
\label{fig:kodak07}
\vspace{-3.0mm}
\end{figure*}

\textbf{Visual Quality.}
Fig. \ref{fig:kodak07} shows the example of reconstructed images (\emph{kodim07.png}) by our methods and the compression standards JEPG, BPG, and VVC (VTM 9.1).
Our reconstructed images retain more details with approximate bpp.
When optimized for MS-SSIM, our CNN-based models and STF models yield significant improvements in visual qualities.

\begin{table}[!t]
\centering
\tabcolsep=7.4pt
\begin{tabular}{@{}lcccc@{}}
\toprule[1.2pt]
Method & Enc (s) & Dec (s) & PSNR$\uparrow$ & bpp $\downarrow$\\
\midrule
Cheng2020 \cite{cheng2020learned} & 8.49 & 14.49 & 35.12 & 0.595\\
Minnen2018 \cite{minnen2018joint} & 16.19 & 21.16 & 35.09 & 0.640\\
\textbf{Our CNN} & \textbf{0.12} & \textbf{0.12} & \textbf{35.91} & \textbf{0.650}\\
\textbf{Our STF} & \textbf{0.15} & \textbf{0.15} & \textbf{35.82} & \textbf{0.651}\\
\bottomrule[1.2pt]
\end{tabular}
\caption{
Comparison of the averaged encoding and decoding time with Minnen2018 \cite{minnen2018joint} and Cheng2020 \cite{cheng2020learned} on Kodak dataset using GPUs (TITAN V).
Note that the results of Cheng2020 are based on a light implementation (without attention module and Gaussian mixture likelihoods) in CompressAI \cite{begaint2020compressai}.
}
\label{tab:efficiency}
\vspace{-3.0mm}
\end{table}

\subsection{Codec Efficiency Analysis}
Spatially Auto-regressive (AR) context model is effective and follow-up studies often use it to enhance the RD performance.
However, the AR model sequentially encodes and decodes each spatial symbol, which significantly slows down the codec efficiency on GPUs and TPUs.
GMM-based entopy model has the same defect. Although it can more precisely estimate PDFs and CDFs of latents, generating CDFs and PDFs dynamically would sacrifice the coding efficiency, while SGM-based entopy model has fixed PDFs and CDFs tables.
Therefore, we use channel-conditional (CC) models \cite{minnen2020channel} as the auto-regressive context model along the channel dimension for better parallel processing, and adopt the SGM-based entopy model for more efficient coding efficiency.

In Table \ref{tab:efficiency}, we evaluate the inference latency of our methods and those time-consuming models \cite{minnen2018joint, cheng2020learned} on the Kodak dataset.

\subsection{Evaluating the Effects of Window Attention}

\begin{figure}[!t]
\centering
\includegraphics[scale=0.5]{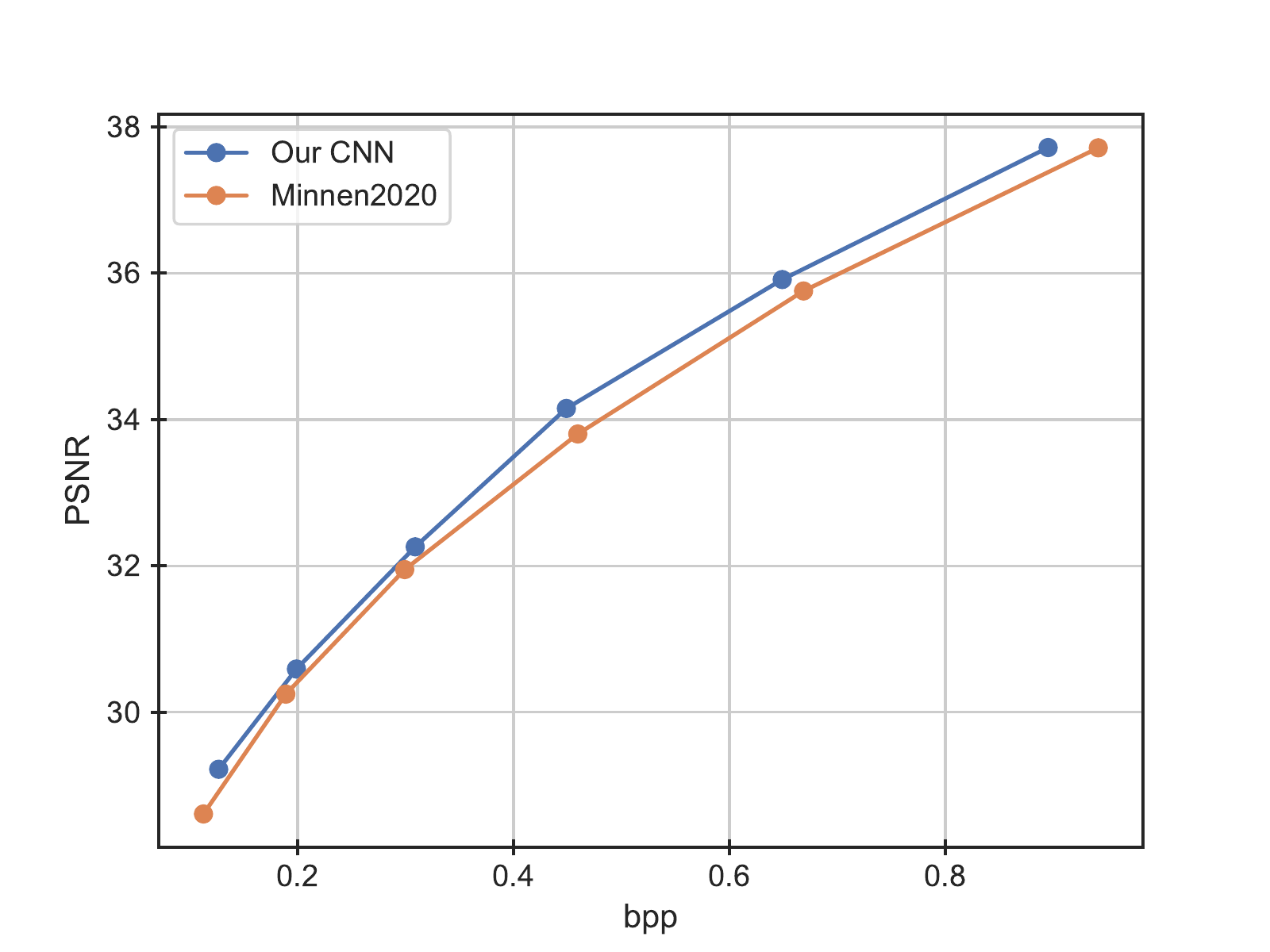}
\caption{Ablation study of window-based attention.
Our model is based on Minnen2020 with WAM.
The RD points of Minnen2020 are obtained from \cite{minnen2020channel}, whose models were trained for 5M steps on the dataset consisting of 2M web images.}
\label{fig:ablation}
\vspace{-3.0mm}
\end{figure}

To prove our conclusion that focusing on spatially neighboring elements can achieve better RD performance, we present contrast experiments by removing the WAMs or using NLAMs, as shown in Table.\ref{tab:ablations}. 
Ablation study in Fig. \ref{fig:ablation} shows that the proposed window-based attention is effective and could enhance current SOTA CNN-based model \cite{minnen2020channel}.

\begin{table}[!t]
\centering
\tabcolsep=12pt
\begin{tabular}{@{}lccc@{}}
\toprule[1.2pt]
Method & $\lambda$ & PSNR $\uparrow$ & bpp $\downarrow$  \\
\midrule
Baseline & \emph{0.0035} & 30.32 & 0.198 \\
Baseline + NLAM & \emph{0.0035} & 30.42 & 0.202 \\
\textbf{Baseline + WAM} & \emph{0.0035} & \textbf{30.58} & \textbf{0.199} \\
\midrule
Baseline & \emph{0.0130} & 33.87 & 0.454\\
Baseline + NLAM & \emph{0.0130} & 33.95 & 0.467\\
\textbf{Baseline + WAM}& \emph{0.0130} & \textbf{34.10} & \textbf{0.448}\\
\bottomrule[1.2pt]
\end{tabular}
\caption{
\textbf{Contrast Experiment.} Baseline is Minnen2020 (SOTA CNN-based model) \cite{minnen2020channel}.
}
\label{tab:ablations}
\vspace{-3.0mm}
\end{table}

\subsection{Discussion}
\textbf{More Efficient Window Attention Designing.} Our CNN-based architecture still have room for improvement, because the window-based attention is not powerful enough for capturing the structural information.
Besides, our experimental results show that GDN and attention mechanism are incompatible.
We guess that GDN is more of a nonlinear activation function than a normalization block.
Our follow-up experiments show that directly using convolution layers to generate attention masks can also achieve comparable RD performance.

\textbf{Compatible Normalization for Transformer.} In our Transformer-based architecture, we use the layer normalization (LN) as default.
The drawback is that LN rescales the responses of linear filters in the network so as to keep it in a reasonable operating range with an identical rescaling factor across all spatial locations, which may destroy the Gaussian distributions of elements.
When computing the attention map, the LN is necessary to rescale the response range.
It seems contradictory but implies that Transformer-based architecture has more potential in learned image compression. Moreover, we are looking forward to combining the Transformer and convolution blocks, our results so far show that they are complementary.

\textbf{Human Perception.} Learned image compression models directly optimize metrics such as PSNR or MS-SSIM and achieve high RD performance.
However, some works\cite{patel2019human,patel2020hierarchical,blau2019rethinking,patel2019deep,patel2021saliency,blau2018rethinking} propose neither of PSNR and MS-SSIM correlate well with human perception.
We found that models optimized for MSE would result in blurred image, and models optimized for MS-SSIM would remove details from regions with high contrast (e.g., text and salient objects), as shown in Fig.\ref{fig:attention contrast}.
\cite{patel2021saliency} proposed a new metric, which is learned on perceptual similarity data specific to image compression.
Inspired by the mathematical definition of perceptual quality in \cite{blau2018rethinking}, \cite{blau2019rethinking} studied the rate-distortion-perception trade-off.
For a more impartial evaluation, perceptual quality measure are vital for practical applications(e.g.,LPIPS \cite{Zhang_2018_CVPR}, FID \cite{heusel2017gans}, KID \cite{binkowski2018demystifying}).

\section{Conclusion}
In this paper, we have extensively studied the local-aware attention mechanism and have found that it is crucial to combine the global structure learned by neural networks and the local texture mined by the attention unit, we have presented a flexible window-based attention module to capture correlations among spatially neighboring elements, which could work as a plug-and-play component to enhance CNN or Transformer models. Furthermore, we have proposed a novel Symmetrical TransFormer (STF) framework with absolute transformer blocks in both down-sampling encoder and up-sampling decoder. Extensive experimental results show that the proposed methods are effective and exceed the state-of-the-art (SOTA) RD performance. In future, we will deeply explore other factors that affect the local detail reconstruction in image compression, such as the convolution kernel shaping and the normalization mode.

\section*{Acknowledgements}
This work was supported in part by the Major Project for New Generation of AI (No.2018AAA0100400) and the National Natural Science Foundation of China (No.61836014, No.U21B2042, No.62006231, No.62072457).
Zou is immensely grateful to the comments from Fan Li that greatly improve the manuscript.

{\small
\bibliographystyle{ieee_fullname}
\bibliography{egbib}
}

\end{document}